\documentclass{emulateapj}
\usepackage{apjfonts}
\journalinfo{The Astrophysical Journal, 2010, in press}

\shorttitle{TANGLED MAGNETIC FIELDS IN PROMINENCES}
\shortauthors{VAN BALLEGOOIJEN \& CRANMER}

\begin{document}

\title{Tangled Magnetic Fields in Solar Prominences}
\author{A. A. van Ballegooijen and S. R. Cranmer}
\affil{Harvard-Smithsonian Center for Astrophysics\\
60 Garden Street, MS-15, Cambridge, MA 02138, USA}

\begin{abstract}
Solar prominences are an important tool for studying the structure and
evolution of the coronal magnetic field. Here we consider so-called
``hedgerow'' prominences, which consist of thin vertical threads.
We explore the possibility that such prominences are supported by
tangled magnetic fields. A variety of different approaches are used.
First, the dynamics of plasma within a tangled field is considered.
We find that the contorted shape of the flux tubes significantly
reduces the flow velocity compared to the supersonic free fall that
would occur in a straight vertical tube. Second, linear force-free
models of tangled fields are developed, and the elastic response of
such fields to gravitational forces is considered. We demonstrate that
the prominence plasma can be supported by the magnetic pressure of a
tangled field that pervades not only the observed dense threads but
also their local surroundings. Tangled fields with field strengths of
about 10 G are able to support prominence threads with observed
hydrogen density of the order of $10^{11}$ $\rm cm^{-3}$. Finally,
we suggest that the observed vertical threads are the result of
Rayleigh-Taylor instability. Simulations of the density distribution
within a prominence thread indicate that the peak density is much
larger than the average density. We conclude that tangled fields
provide a viable mechanism for magnetic support of hedgerow
prominences.
\end{abstract}


\section{Introduction}
\label{sect:intro}

\begin{figure}
\epsscale{1.13}
\plotone{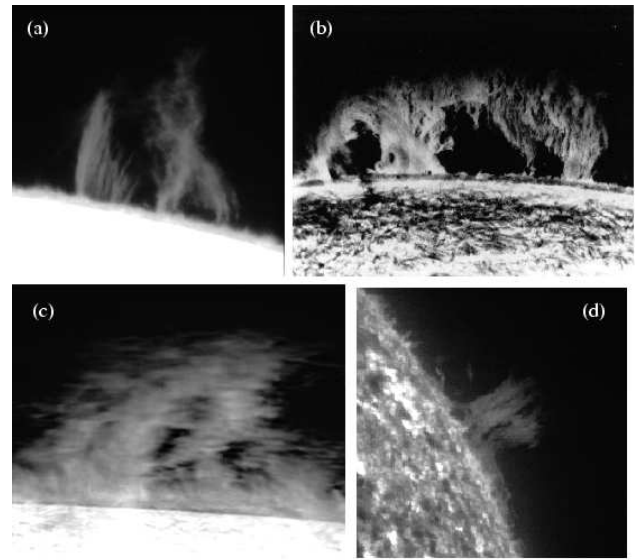}
\caption{Quiescent prominences above the solar limb: (a) H$\alpha$
image of prominence observed at Big Bear Solar Observatory (BBSO),
November 22, 1995; (b) H$\alpha$, BBSO, 1970; (c) H$\alpha$, Dutch
Open Telescope, September 15, 2006; (d) He~II 304 {\AA}, STEREO/EUVI,
2008 April 20 at 00:06 UT.}
\label{fig:proms}
\end{figure}

Solar prominences (a.k.a.~filaments) are relatively cool structures
embedded in the million-degree corona at heights well above the
chromosphere \citep[see reviews by][] {Hirayama1985, Zirker1989,
Priest1990, Tan-Han1995, Heinzel2007}. Above the solar limb,
prominences appear as bright structures against the dark background,
but when viewed as filaments on the solar disk they can be brighter or
darker than their surroundings, depending on the bandpass used to
observe them. Magnetic fields are thought to play an important role in
supporting the prominence plasma against gravity, and in insulating it
from the surrounding hot corona. Most quiescent prominences exhibit
intricate filamentary structures that evolve continually due to plasma
flows and heating/cooling processes \citep[see examples in][]
{Menzel1960, Engvold1976, Malherbe1989, Leroy1989, Martin1998}.
In some cases the threads appear to be mostly horizontal, while in
other cases they are clearly radially oriented (``hedgerow''
prominences). Figure \ref{fig:proms} shows several examples of
prominences observed in H$\alpha$ at the Big Bear Solar Observatory
(BBSO) and the Dutch Open Telescope (DOT). The examples in
Figs.~\ref{fig:proms}a and \ref{fig:proms}b show mainly vertical
threads, while the prominence in Fig.~\ref{fig:proms}c shows
horizontal threads. Off-limb observations in He~II 304 {\AA} indicate
that there are higher altitude parts that are optically thin in
H$\alpha$ and therefore not visible on the disk (or at least have not
been clearly identified in disk observations).  Figure
\ref{fig:proms}d shows a prominence above the solar limb as observed
in He~II 304 {\AA} with the SECCHI/EUVI instrument \citep[][]
{Howard2007} on the STEREO spacecraft. The upper parts of the
prominence consist of vertical threads with an intricate fine-scale
structure. Movie sequences of quiescent and erupting prominences can
be found at the STEREO
website\footnote{http://stereo.gsfc.nasa.gov/gallery/selects.shtml}.

Prominence plasma is highly dynamic, exhibiting horizontal and
vertical motions of order 10--70 km~s$^{-1}$ \citep{Menzel1960,
Engvold1976, Zirker_etal1998, Kucera2003, Lin2003, Okamoto2007,
Berger2008, Chae2008}. 
Recent high-resolution observations of filaments on the solar disk
indicate that they consist of a collection of very thin threads with
widths of about 200 km, at the limit of resolution of present-day
telescopes \citep[][]{Lin2003, Lin2005a, Lin2005b, Lin2008a,
Lin2008b}. Individual threads have lifetimes of only a few minutes,
but the filament as a whole can live for many days.  It seems likely
that these thin threads are aligned with the local magnetic field.
High-resolution images of prominences above the limb have been
obtained with the Solar Optical Telescope (SOT) onboard Hinode.
For example, \citet{Okamoto2007} observed horizontal threads in a
prominence near an active region, and studied the oscillatory motions
of these threads. \citet{Heinzel2008} observed a hedgerow prominence
consisting of many thin vertical threads, and they used
multi-wavelength observations to estimate the amount of absorption and
``emissivity blocking'' in the prominence and surrounding cavity.
\citet{Berger2008} observed another hedgerow prominence and found
that the prominence sheet is structured by both bright quasi-vertical
threads and dark inclusions. The bright structures are downflow
streams with velocity of about 10 km~s$^{-1}$, and the dark inclusions are
highly dynamic upflows with velocity of about 20 km~s$^{-1}$. The downflow
velocities are much less than the free-fall speed, indicating that the
plasma is somehow being supported against gravity. \citet{Berger2008}
proposed that the dark plumes contain relatively hot plasma that is
driven upward by buoyancy. \citet{Chae2008} observed a persistent flow
of H$\alpha$ emitting plasma into a prominence from one side, leading
to the formation of vertical threads. They suggested that the vertical
threads are stacks of plasma supported against gravity by the sagging
of initially horizontal magnetic field lines.

Direct measurements of the prominence magnetic field can be obtained
using spectro-polarimetry \citep[see reviews by][] {Leroy1989,
Paletou2003, Paletou2008, Lopez2007}. A comprehensive effort to
measure prominence magnetic fields was conducted in the 1970's and
early 1980's using the facilities at Pic du Midi (France) and
Sacramento Peak Observatory (USA). This work showed that (1) the
magnetic field in quiescent prominences has a strength of 3--15 G;
(2) the field is mostly {\it horizontal} and makes an angle of about
$40^\circ$ with respect to the long axis of the prominence \citep[][]
{Leroy1989, Bommier1998, Paletou2003}; (3) the field strength
increases slightly with height, indicating the presence of dipped
field lines; (4) most prominences have {\it inverse} polarity, i.e.,
the component of magnetic field perpendicular to the prominence axis
has a direction opposite to that of the potential field. These earlier
data likely included a variety of quiescent prominences, some with
predominantly horizontal threads, others with more vertical threads.
In more recent work, \citet{Paletou2001} reported full-Stokes
observations of a limb prominence in He~I 5876 {\AA} (He~$\rm D_3$),
and derived magnetic field strengths of 30--45 G, somewhat larger than
the values reported in earlier studies. \citet{Casini2003} published
the first vector-field map of a prominence with a spatial resolution
of a few arcseconds. They found that the average magnetic field in
this prominence is mostly horizontal with a strength of about 20 G
and with the magnetic vector pointing $20^\circ$ to $30^\circ$ off
the prominence axis, consistent with the earlier studies. However,
the map also shows clearly organized patches where the magnetic field
is significantly stronger than average, up to 80 G \citep[also see][]
{Wiehr2003, Lopez2002, Lopez2003, Lopez2007, Casini2005}. It is
unclear how these patches are related to the fine threads seen at
higher spatial resolution. Recently, \citet{Merenda2006} observed
He~I 10830 {\AA} in a polar crown prominence above the limb, and
found evidence for fields of about 30 G that are oriented only
$25^\circ$ from the vertical direction.

How is the plasma in hedgerow prominences supported against gravity?
Many authors have suggested that quiescent prominences are embedded in
large-scale {\it flux ropes} that lie horizontally above the polarity
inversion line \citep[][] {Kuperus1974, Pneuman1983, vanB1989,
Priest1989, Rust1994, Low1995, Aulanier1998, Chae2001, Gibson2006,
Dudik2008}. The prominence plasma is thought to be located near the
{\it dips} of the helical field lines. The magnetic field near the
dips may be deformed by the weight of the prominence plasma \citep[][]
{Kippenhahn1957, Low2005, Petrie2005, Heinzel2005}. Others have
suggested that the magnetic field in hedgerow prominences is vertical
along the observed threads, and that the plasma is supported by MHD
waves \citep[][] {Jensen1983, Jensen1986, Pecseli2000}. However,
relatively high frequencies and wave amplitudes are required, and it
is unclear why such waves would not lead to strong heating of the
prominence plasma. \citet{Dahlburg1998} and \citet{Antiochos1999}
showed that the prominence plasma can be supported by the pressure of
a coronal plasma lower down along an inclined field line; however,
this mechanism only works for nearly horizontal field lines
\citep[also see][] {Mackay2001, Karpen2005, Karpen2006, Karpen2008}.
Furthermore, hedgerow prominences are located in coronal cavities
where the plasma pressure is very low.  Therefore, it seems unlikely
that the vertical threads seen in hedgerow prominences can be
supported by coronal plasma pressure on nearly vertical field lines.

In this paper we propose that hedgerow prominences are embedded in
magnetic fields with a complex ``tangled'' structure. Such tangled
fields have many dips in the field lines where the weight of the
prominence plasma can be counteracted by upward magnetic forces.
Our purpose is to demonstrate that such tangled fields provide a
viable mechanism for prominence support in hedgerow prominences.
\citet{Casini2009} recently invoked tangled fields in the
interpretation of spectropolarimetric observations of an active
region filament. While such filaments are quite different from the
hedgerow prominences considered here, this work shows that tangled
fields have important effects on the measurement of prominence
magnetic fields. Such effects will not be considered in this paper.

The paper is organized as follows. In Section 2 we propose that
hedgerow prominences are supported by tangled magnetic fields, and we
discuss how such fields may be formed. In Section 3 we present a
simple model for the dynamics of plasma along the tangled field lines,
and we show that weak shock waves naturally occur in such plasmas. In
Section 4 we develop a magnetostatic model of tangled fields based on
the {\it linear force-free field} approximation, and in Section 5 we
study the response of such fields to gravitational forces. In Section
6 we simulate the distribution of plasma in a cylindrical prominence
thread. In Section 7 we discuss the formation of vertical threads by
Rayleigh-Taylor instability. The results of the investigation are
summarized and discussed in Section 8.

\section{Tangled Fields in Prominences}
\label{sect:tangle}

\begin{figure}
\epsscale{1.15}
\plotone{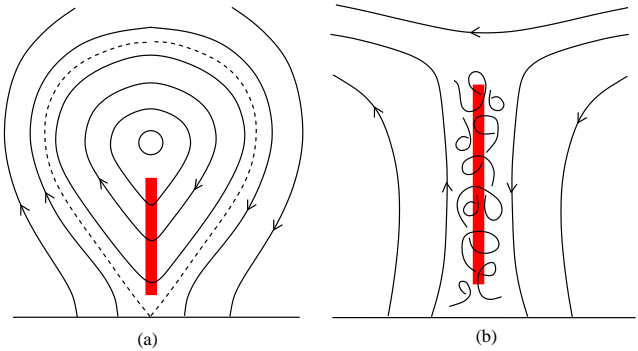}
\caption{Magnetic support of solar prominences: (a) by a large-scale
coronal flux rope, (b) by a tangled magnetic field in a current
sheet.}
\label{fig:cartoon2}
\end{figure}

The spectro-polarimetric observations of prominences described in
Section~\ref{sect:intro} are consistent with the idea that quiescent
prominences are embedded in coronal flux ropes that lie horizontally
above the polarity inversion line (PIL). Figure~\ref{fig:cartoon2}a
shows a vertical cross-section through such a flux rope. The magnetic
field also has a component into the image plane, so the field lines
are helices, and the plasma is assumed to be located at the dips of
these helical windings. A dip is defined as a point where the magnetic
field is locally horizontal and curved upward. As indicated in the
figure, the magnetic field may be deformed by the weight of the
prominence plasma, creating V-shaped dips. The magnetic field near
such dips is well described by the Kippenhahn-Schl\"{u}ter model
\citep[][] {Kippenhahn1957}, and several authors have developed local
magnetostatic models of the fine structures observed in quiescent
prominences \citep[e.g.,][]{Low1982, Petrie2005, Heinzel2005}.
However, recent observations of ``dark plumes'' \citep[][]
{Berger2008} and rotational motions \citep[][]{Chae2008} within
prominences remind us again that prominences have complex internal
motions, and it is not clear how such motions can be explained in
terms of a single large flux rope. Perhaps the magnetic structure of
hedgerow prominences is more complicated than that predicted by the
flux rope model (Figure~\ref{fig:cartoon2}a).

In this paper we propose an alternative model, which is illustrated
in Figure~\ref{fig:cartoon2}b. Following \citet{Kuperus1967}, we
suggest that hedgerow prominences are formed in current sheets that
overlie certain sections of the PIL on the quiet Sun. Unlike those
previous authors we suggest that the current sheet extends only to
limited height ($\sim 100$ Mm), and may extend only a limited distance
along the PIL. Furthermore, we propose that {\it tangled} magnetic
fields are present within these current sheets. A tangled field is
defined as a magnetic structure in which the field lines are woven
into an intricate fabric, and individual field lines follow nearly
random paths. We suggest that the field is tangled on a spatial scale
of 0.1--1 Mm, comparable to the pressure scale height $H_p$ of the
prominence plasma ($H_p \approx 0.2$ Mm). The prominence plasma is
assumed to be located at the many dips of the tangled field lines.
The tangled field is confined horizontally by the vertical fields on
either side of the sheet, and vertically by the weight of the
prominence plasma.

A key feature of a tangled field is that the plasma and field are in
magnetostatic equilibrium, i.e., the Lorentz force is balanced by
the gas pressure gradients and gravity.  Therefore, a tangled field
is quite different from ``turbulent'' magnetic fields in which
large-amplitude Alfv\'{e}n waves are present (e.g., the solar wind).
In a tangled field the magnetic perturbations do not propagate along
the field lines. In this paper we examine the basic properties of
tangled fields, and we investigate their ability to support the
prominence plasma.

We suggest that the tangled field may be formed as a result of
magnetic reconnection, not the twisting or stressing of field lines.
Quiescent prominences are located above polarity inversion ``lines''
that are often more like wide bands of mixed polarity separating
regions with dominantly positive and negative polarity. In these
mixed-polarity zones, magnetic flux elements move about randomly and
opposite polarity elements may cancel each other \citep[e.g.,][]
{Livi1985}. New magnetic bipoles frequently emerge from below the
photosphere. These processes causes the ``recycling'' of the
photospheric flux about once every 2 to 20 hours \citep[][]
{Hagenaar2003, Hagenaar2008}, and the coronal flux is recycled even
faster \citep[][]{Close2005}. It is likely that the interactions
between these flux elements produce a complex, non-potential magnetic
field in the low corona. Within this environment magnetic reconnection
is likely to occur frequently at many different sites in the corona
above the inversion zone. Each reconnection event may produce a bundle
of twisted field, and the twisted fields from different events may
collect into larger conglomerates to form a tangled field. The
tangled field may rise to larger heights (as a result of its natural
buoyancy), and may collect into a thick sheet that is sandwiched
between smoother fields, as illustrated in Figure
\ref{fig:cartoon2}b. The observed prominence consists of plasma that
is trapped within this sheet. New tangled field is continually
injected into the sheet from below, producing vertical motions within
the sheet. We suggest that the ``dark plumes'' observed by
\citet{Berger2008} may be a manifestation of such vertical motions of
the tangled field.

\section{Flows Along the Tangled Field}
\label{sect:flows}

\begin{figure}
\epsscale{0.41}
\plotone{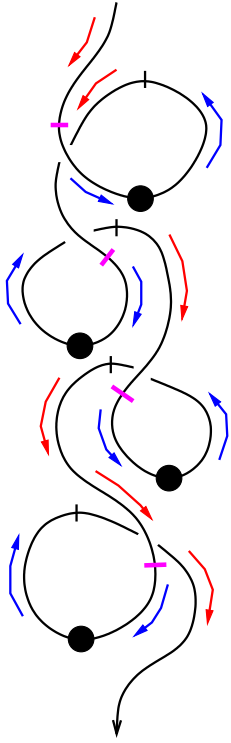}
\caption{Flows along a highly distorted field line in the tangled
magnetic field of a solar prominence. The field line is indicated by
the {\it solid curve}, and the {\it black dots} indicate ``dips'' in
the field line where cool prominence plasma can collect. The arrows
show the direction of subsonic ({\it blue}) and supersonic ({\it red})
flows. Sonic points are located at peaks in the field lines ({\it
black vertical bars}), and shock waves occur where supersonic flows
slow down before reaching a dip ({\it magenta bars}).}
\label{fig:flow_cartoon}
\end{figure}

The spatial distribution of plasma within the prominence is determined
in part by the dynamics of plasma along the tangled field lines.
Figure \ref{fig:flow_cartoon} shows the contorted (but generally downward)
path of an individual field line in the tangled field. Note that there
are several ``dips'' where the field line is horizontal and curved
{\it upward}, and ``peaks'' where the field is horizontal and curved
{\it downward}.  Tracing the field line upward from a dip, one always
reaches a peak where the field line again turns downward. Therefore,
the question arises whether the plasma collected in the dips would
remain in these dips or be siphoned out of the dips via the peaks of
the field lines.

To answer this question, we consider a simple model for the motion of
the prominence plasma along the magnetic field. For simplicity we
assume that the flow takes place in a thin tube surrounding the
selected field line (i.e., the divergence of neighboring field lines
is neglected), and the cross-sectional area of this tube is taken to
be constant. We assume a steady flow is established along the tube.
Let $v(s)$ and $\rho (s)$ be the plasma velocity and density as
functions of position $s$ along the tube, then conservation of mass
requires $\rho v$ = constant. The equation of motion of the plasma is
\begin{equation}
\rho v \frac{dv}{ds} = - \frac{dp}{ds} - \rho g \frac{dz}{ds} ,
\label{eq:flow1}
\end{equation}
where $p(s)$ is the plasma pressure, $z(s)$ is the height above the
photosphere, and $g$ is the acceleration of gravity. The equation of
state is written in the form $p = K \rho^\gamma$, where $\gamma$ and
$K$ are constants (we use $\gamma < 5/3$ to describe non-adiabatic
processes). Eliminating $p(s)$ and $\rho (s)$ from equation
(\ref{eq:flow1}), we obtain the following equation for the parallel
flow velocity:
\begin{equation}
\left( v - \frac{c^2}{v} \right) \frac{dv}{ds} = - g \frac{dz}{ds} ,
\label{eq:flow2}
\end{equation}
where $c(s)$ is the sound speed ($c^2 \equiv \gamma p/\rho$). The
above equation has a critical point where the flow velocity equals
the sound speed ($v = c$). Therefore, a transition from subsonic to
supersonic flow can occur only at points where the RHS of this
equation vanishes, $dz/ds = 0$. These sonic points are located at the
peaks of the field lines where matter can be siphoned out of one dip
and deposited into another dip at lower height. The resulting flow
pattern is indicated in Figure \ref{fig:flow_cartoon}. As the
supersonic flow approaches the next dip, it must slow down to subsonic
speeds, which can only happen in a shock. Therefore, the tube has a
series of subsonic and supersonic flows separated by shocks and sonic
points. The role of these shocks is to dissipate the gravitational
energy that is released by the falling matter.

The position and strength of the shocks can be computed if the height
$z(s)$ of the flow tube is known. Neighboring peaks are generally not
at the same height. Therefore, each section between neighboring peaks
is approximated as a large-amplitude sinusoidal perturbation
superposed on a generally downward path:
\begin{equation}
z(s) \approx A \cos \left( 2 \pi \frac{s}{\Lambda} - \phi_0 \right)
- C s , \label{eq:zsapp}
\end{equation}
where $s$ is the position along the flow tube, $\Lambda$ is the
distance between neighboring peaks (as measured along the flow tube),
$A$ is the amplitude of the perturbation in height, $\phi_0$ is a
phase angle, and $C$ is the background slope. The phase angle is
chosen such that the peaks in the flow tube (where $dz/ds = 0$) are
located at $s = 0$ and $s = \Lambda$, then the slope is given by
\begin{equation}
C = 2 \pi \frac{A} {\Lambda} \sin \phi_0 .
\end{equation}
The sonic points will then be located at $s = 0$ and $s = \Lambda$.
\begin{figure}
\epsscale{1.13}
\plotone{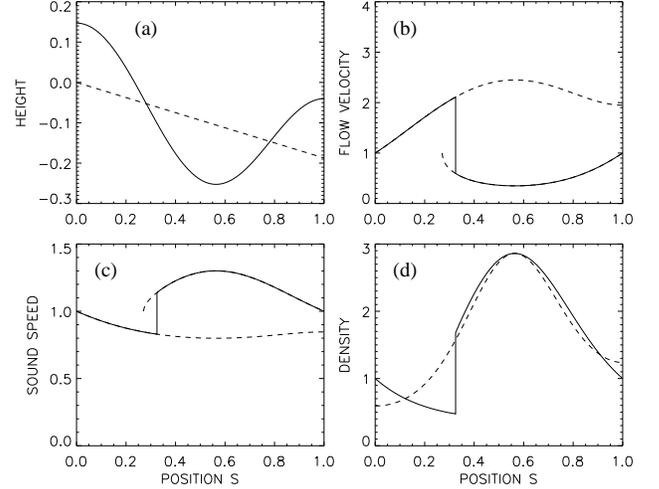}
\caption{Model for plasma flow along a single field line in a tangled field.
(a) The {\it solid} curve shows the height $z(s)$ as function of
position $s$ along the field line [in units of the distance $\Lambda$
between neighboring peaks, see equation (\ref{eq:zsapp})]. The {\it
dashed} curve shows the overall downward trend of the flow tube. (b)
The {\it solid} curve shows the parallel flow velocity $v(s)$ (in
units of $c_0$) for $\gamma = 1.5$ and $H_{p,0} = 0.15 \Lambda$. The
{\it dashed} curves show the subsonic and supersonic solutions of
equation (\ref{eq:flow3}). The sonic points are located at the
peaks in the field line ($s=0$ and $s=\Lambda$), and the shock is
located at $s = 0.325 \Lambda$. (c) Similar plot for the sound speed
$c(s)$ (in units of $c_0$). (d) The {\it solid} curve shows the plasma
density $\rho (s)$ along the tube (in arbitrary units). The
hydrostatic equilibrium density is shown by the {\it dashed} curve.}
\label{fig:flow_shock}
\end{figure}
Figure \ref{fig:flow_shock}a shows the height $z(s)$ for $A = 0.15
\Lambda$ and $\phi_0 = 0.2$ rad, so that $C = 0.187$. Let $p_0$,
$\rho_0$ and $c_0$ be the pressure, density and sound speed at the
sonic points, then $K = \gamma^{-1} c_0^2 \rho_0^{1-\gamma}$, and the
sound speed can be written as
\begin{equation}
c(s) = c_0 [ \rho(s) / \rho_0 ]^{(\gamma-1)/2}
= c_0 [ v(s)/c_0 ]^{-(\gamma-1)/2} .
\end{equation}
Inserting this expression into equation
(\ref{eq:flow2}), we obtain
\begin{equation}
\left( u - u^{-\gamma} \right) \frac{du}{ds} = - \frac{1}
{\gamma H_{p,0}} \frac{dz}{ds} , \label{eq:flow3}
\end{equation}
where $u(s) \equiv v(s)/c_0$, and $H_{p,0} \equiv p_0/(\rho_0 g)$ is
the pressure scale height at the sonic points.
Equation (\ref{eq:flow3}) can be integrated as follows:
\begin{equation}
\onehalf [ u^2(s) - 1 ] + \frac{1}{\gamma-1} \left[ u^{1-\gamma}(s)
- 1 \right] = - \frac{z(s)-z(s_0)} {\gamma H_{p,0}} ,
\label{eq:flow4}
\end{equation}
where $s_0$ is the position of a sonic point. Equation
(\ref{eq:flow4}) can be solved for $u(s)$ by Newton-Raphson iteration.
When $\gamma = 1$, there is an analytic solution for $u(s)$ in terms
of the Lambert $W$ function \citep[see][]{Cranmer2004}; in the present
paper we assume $\gamma = 1.5$.
The supersonic solution $v_1(s)$ is computed with $s_0 = 0$, and the
subsonic solution $v_2(s)$ is computed with $s_0 = \Lambda$. The
dashed curves in Figure \ref{fig:flow_shock}b show $v_1(s)$ and
$v_2(s)$, and Figure \ref{fig:flow_shock}c shows the corresponding
sound speeds $c_1 (s)$ and $c_2(s)$. Here we assumed a scale height
$H_{p,0} = 0.15 \Lambda$, equal to the amplitude of the field-line
distortions. The Mach number of the flow is given by
\begin{equation}
M(s) = v(s)/c(s) = [ u(s) ]^{(\gamma+1)/2} .
\end{equation}
The shock is located at the point $s_{sh}$ where the Mach number $M_1$
before the shock and the Mach number $M_2$ after the shock satisfy the
following relationship:
\begin{equation}
M_2^2 = \frac{2 + (\gamma-1) M_1^2} {2 \gamma M_1^2 - (\gamma-1)} ,
\end{equation}
which follows from the Rankine-Hugionot conditions for parallel shocks
\citep[][]{Landau1959}. Therefore, the actual flow velocity $v(s)$
between the two sonic points is given by the full curve in Figure
\ref{fig:flow_shock}b, and the sound speed $c(s)$ is given by the full
curve in Figure \ref{fig:flow_shock}c.

The plasma density $\rho (s)$ along the flow tube is determined by
mass conservation ($\rho v$ = constant), and is plotted in Figure
\ref{fig:flow_shock}d. Note that there is a strong peak in the density
at the dip in the field line, $s_{dip} \approx 0.56 \Lambda$. The dashed
curve in Figure \ref{fig:flow_shock}d shows the density profile that
would exist if the plasma were in hydrostatic equilibrium (HE):
\begin{equation}
\rho_{HE} (s) = \rho_{dip} \exp \left[ - \frac{z(s) - z(s_{dip})}
{H_p} \right] , \label{eq:rho}
\end{equation}
where $\rho_{dip}$ and $H_p$ are the density and pressure scale
height at the dip, $H_p = 0.254 \Lambda$. The deviations from
hydrostatic equilibrium are significant only in those regions where
the flow velocity is comparable to the sound speed. We define an
average flow velocity $\bar{v}$ by
\begin{equation}
\bar{v} \equiv \frac{\int_0^{\Lambda} \rho(s) v(s) ds}
{\int_0^{\Lambda} \rho (s) ds} = \frac{\Lambda} {\int_0^{\Lambda}
v^{-1} (s) ds} .
\end{equation}
For the case shown in Figure \ref{fig:flow_shock} we find $\bar{v} =
0.6 c_0$, so the average flow speed is less than the sound speed.
Therefore, the contorted shape of the flow tube significantly reduces
the flow velocity compared to the supersonic free fall that would
occur in a straight vertical tube.

The cooler parts of the prominence are thought to have a temperature
$T \sim 10^4$ K. Assuming a hydrogen ionization fraction of 10\%,
a helium abundance of 10 \% and $\gamma = 1.5$, the sound speed
$c_0 \approx 10$ km~s$^{-1}$, and we predict an average flow velocity
$\bar{v} \approx 0.6 c_0 \approx 6$ km~s$^{-1}$.
The vertical component of this velocity is $\bar{v}_z \approx
- C \bar{v} \approx - 1.1$ km~s$^{-1}$,
less than the observed vertical velocities in prominence threads
(5--10 km~s$^{-1}$). Note that the predicted velocity is
relative to the {\it pattern} of the tangled field, therefore, if the
tangled field expands in the vertical direction it will push the
prominence plasma upward.  We speculate that the observed upward
motions in hedgerow prominences \citep[e.g.,][]{Berger2008} are due
to such large-scale changes in the tangled field.

\section{Linear Force-Free Field Models}
\label{sect:LFFF}

In this section simple models for tangled fields are developed.
A volume $V$ in the corona is considered, and the plasma inside this
volume is assumed to be in magnetostatic equilibrium, $- \nabla p
+ \rho {\bf g} + {\bf F} = 0$, where $p$ is the plasma pressure,
$\rho$ is the density, ${\bf g}$ is the acceleration of gravity,
and ${\bf F}$ is the Lorentz force. All quantities are functions of
position ${\bf r}$ within the volume. The Lorentz force is given by
\begin{equation}
{\bf F} \equiv \frac{1}{c} {\bf j} \times {\bf B} = \frac{1}{4 \pi}
( \nabla \times {\bf B} ) \times {\bf B} , \label{eq:F}
\end{equation}
where ${\bf j}$ is the electric current density and ${\bf B}$ is
the magnetic field. Using tensor notation, equation (\ref{eq:F}) can
also be written as $F_i = \partial T_{ij} / \partial x_j$, where
$T_{ij}$ is the magnetic stress tensor, a special case of Maxwell's
stress tensor \citep[][]{Jackson1999}:
\begin{equation}
T_{ij} \equiv - \frac{B^2}{8 \pi} \delta_{ij} + \frac{B_i B_j}
{4 \pi} . \label{eq:T}
\end{equation}
The first term describes magnetic pressure, and the second term
describes magnetic tension. In a tangled field both pressure and
tension forces are important.

If gravity and plasma pressure gradients are neglected, then ${\bf F}
\approx 0$, so the magnetic field ${\bf B} ({\bf r})$ must satisfy the
force-free condition:
\begin{equation}
\nabla \times {\bf B} = \alpha {\bf B} , \label{eq:FFF}
\end{equation}
where $\alpha ({\bf r})$ may in general be a function of position.
In the special case that $\alpha$ is constant throughout the volume,
equation (\ref{eq:FFF}) becomes a linear equation for ${\bf B}
({\bf r})$, and the solutions are called linear force free fields
(LFFF). \citet{woltjer1958} has shown that in a closed magnetic system
with a prescribed magnetic helicity ($H \equiv \int {\bf A} \cdot {\bf
B} dV$, where ${\bf A}$ is the vector potential) the lowest-energy
state is a LFFF. Therefore, in this paper only LFFFs are considered,
and $\alpha$ is treated as a free parameter. We find that LFFFs can be
tangled. The typical length scale of the tangled field is given by the
inverse of the $\alpha$ parameter, $\ell \equiv |\alpha|^{-1}$.  In
the following we first consider the case that $\ell$ is small compared
to the domain size $L$ in {\it any} direction, and then consider
the boundary effects. Section \ref{sect:cylinder} describes tangled
fields in a cylindrical domain.

\subsection{Tangled Field in a Large Volume}
\label{sect:volume}

In the absense of boundary conditions, the solution of equation
(\ref{eq:FFF}) can be written as a superposition of planar modes:
\begin{equation}
{\bf B} ({\bf r}) = \sum_{n=1}^N B_n \left[
\hat{\bf e}_{1,n} \cos ( {\bf k}_n \cdot {\bf r} + \beta_n ) -
\hat{\bf e}_{2,n} \sin ( {\bf k}_n \cdot {\bf r} + \beta_n ) \right] ,
\label{eq:sum}
\end{equation}
where $N$ is the number of modes, ${\bf k}_n \equiv \alpha
\hat{\bf e}_{3,n}$ is the wave vector ($n = 1, \cdots, N$),
$B_n$ is the mode amplitude, $\beta_n$ is a phase angle, and
$[\hat{\bf e}_{1,n}, \hat{\bf e}_{2,n}, \hat{\bf e}_{3,n}]$ are
unit vectors that are mutually orthogonal and form a right-handed
basis system:
\begin{eqnarray}
\hat{\bf e}_{1,n} & = & \cos \theta_n  ( \cos \phi_n ~
\hat{\bf y} + \sin \phi_n ~ \hat{\bf z} ) - \sin \theta_n ~
\hat{\bf x} , \label{eq:e1} \\
\hat{\bf e}_{2,n} & = & - \sin \phi_n ~ \hat{\bf y} +
\cos \phi_n ~ \hat{\bf z} , \label{eq:e2} \\
\hat{\bf e}_{3,n} & = & \sin \theta_n  ( \cos \phi_n ~
\hat{\bf y} + \sin \phi_n ~ \hat{\bf z} ) + \cos \theta_n ~
\hat{\bf x} . \label{eq:e3} 
\end{eqnarray}
Here $\theta_n$ and $\phi_n$ are the direction angles of the wave
vector relative to the Cartesian reference frame.

\begin{figure}
\epsscale{1.10}
\plotone{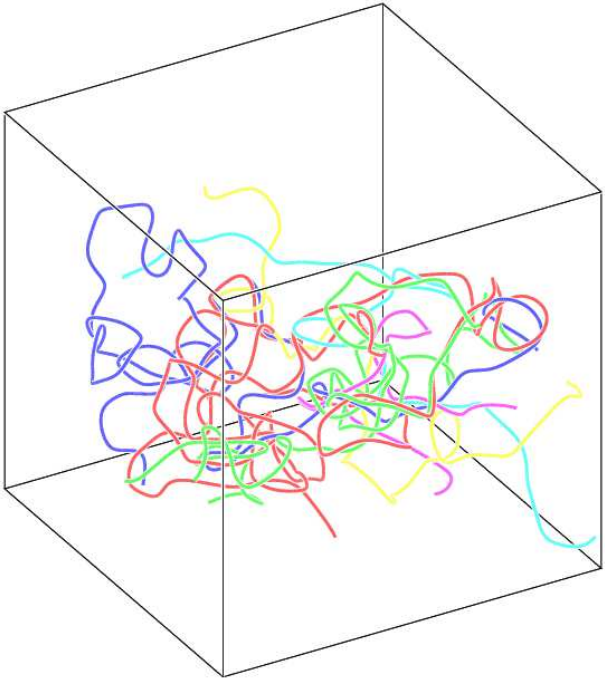}
\caption{Tangled field obtained by superposition of 100 randomly
selected modes of the linear force-free field in a large volume.}
\label{fig:volume}
\end{figure}

Figure \ref{fig:volume} shows an example of a field with $N = 100$
modes, an isotropic distribution of direction angles $(\theta_n,
\phi_n)$, and randomly selected phase angles $\beta_n$. The starting
points of the field lines are randomly selected from the central part
of the box, and the field lines are traced until they reach the box
walls. Note that individual field lines follow random paths, and that
different field lines are tangled together.

Now consider an ensemble ${\cal E}_{\beta,N}$ of fields with different
(randomly distributed) phase angles $\beta_n$, but with a fixed number
of modes $N$, and with fixed mode amplitudes $B_n$ and direction
angles $(\theta_n, \phi_n)$. The phase angles are assumed to be
uniformly distributed in the range $[0, 2 \pi]$, and angles from
different modes $n$ and $n^\prime$ are assumed to be uncorrelated.
Then the ensemble average of the magnetic field vanishes,
\begin{equation}
< {\bf B} >_\beta = 0 , \label{eq:bavg}
\end{equation}
where $< \cdots >_\beta$ denote the average over phase angles
$\beta_n$. Also, the average of the tensor ${\bf B} {\bf B}$ is
given by
\begin{displaymath}
< {\bf B} {\bf B} >_\beta \,\, = \, \onehalf \sum_{n=1}^N B_n^2 \left(
\hat{\bf e}_{1,n} \hat{\bf e}_{1,n} + \hat{\bf e}_{2,n}
\hat{\bf e}_{2,n} \right)
\end{displaymath}
\begin{equation}
 \,\,\,\,\, = \, \onehalf \sum_{n=1}^N B_n^2 \left(
{\bf I} - \hat{\bf e}_{3,n} \hat{\bf e}_{3,n} \right) ,
\label{eq:tens1}
\end{equation}
where ${\bf I}$ is the unit tensor (${\bf I} = \hat{\bf x} \hat{\bf x}
+ \hat{\bf y} \hat{\bf y} + \hat{\bf z} \hat{\bf z}$). Expression
(\ref{eq:tens1}) is independent of position ${\bf r}$, so the magnetic
field is statistically {\it homogeneous}, but is not necessarily
isotropic. 

We study the statistical properties of field lines in models with
different values of the mode number $N$. For each $N$ we construct a
series of models ($m = 1, \cdots, M$) with different phase angles,
but with constant values of the mode amplitudes $B_n$ and direction
angles $(\theta_n,\phi_n)$. The mode amplitudes are taken to be the
same for all modes ($B_n = 1$). For each realization $m$ of the
phase angles $\beta_n$ we trace out the field line that starts at
the origin ($x = y = z = 0$), and we measure the square of the radial
distance $r_m^2 (s)$ as function of position $s$ along the field line:
\begin{equation}
r_m^2 (s) = x_m^2(s) + y_m^2(s) + z_m^2(s) .
\end{equation} 
We then average this quantity over $M = 100$ realizations of the phase
angles to obtain the mean square distance $\overline{r^2} (s)$.
For $N = 3$ both mutually orthogonal directions $(\theta_n,\phi_n)$
and randomly chosen directions are considered. In both cases we find
that the field lines follow long helical paths, and $\overline{r^2}
(s)$ increases quadratically with $s$. Therefore, for $N = 3$ the
field lines do not behave randomly. For $N \ge 4$ only randomly chosen
directions are considered. For $N = 4$ some of the field lines are
long helices, while others have more random paths, and for $N=5$ all
field lines seem random, however, in both cases $\overline{r^2} (s)$
is not well fit by a power law. True random walk behavior of the field
lines, as indicated by a {\it linear} dependence of $\overline{r^2}$
on distance $s$, is found only when the number of modes is increased
to $N \ge 10$. In the limit of large $N$, $\overline{r^2} (s) \approx
10 s / |\alpha|$.

We now consider a larger ensemble ${\cal E}_N$ in which not only the
phase angles $\beta_n$ but also the mode amplitudes $B_n$ and
direction angles $(\theta_n, \phi_n)$ are allowed to vary. From now on
$< \cdots >$ will denote an average over this larger ensemble.
The direction angles are assumed to have an isotropic distribution,
i.e., the angle $\phi_n$ is uniformly distributed in the range
$[0,2\pi]$, and $\cos \theta_n$ is uniform in the range $[-1,+1]$,
so that $< \cos^2 \theta_n > = \onethird$. Furthermore, the mode
amplitudes $B_n$ are assumed to be uncorrelated with the direction
angles, and $< B_n^2 > = B_0^2 /N$, where $B_0$ is a constant.
Then $< {\bf B} > = 0$, so the mean magnetic field vanishes.
Further averaging of equation (\ref{eq:tens1}) shows that
$< {\bf B} {\bf B} >$ is an isotropic tensor:
\begin{equation}
< {\bf B} {\bf B} > = \onehalf \sum_{n=1}^N < B_n^2 > \left(
{\bf I} - < \hat{\bf e}_{3,n} \hat{\bf e}_{3,n} > \right)
= \onethird B_0^2 {\bf I} . \label{eq:tens2} 
\end{equation}
It follows that $< B^2 > = B_0^2$, so $B_0$ equals the r.m.s.~value of
the total field strength. The ensemble average of the magnetic stress
tensor, equation (\ref{eq:T}), is given by
\begin{equation}
< T_{ij} > = - \frac{B_0^2} {24 \pi} \delta_{ij} , \label{eq:stress}
\end{equation}
which is also isotropic. Note that the diagonal components of
$< T_{ij} >$ are negative, so the effects of magnetic pressure
dominate over the effects of magnetic tension. Therefore, the
isotropic tangled field has a positive magnetic pressure,
$p_t = B_0^2 / (24 \pi)$. The average energy density is $E_t =
B_0^2 / (8 \pi)$. The relationship $E_t = 3 p_t$ is similar to that
for a relativistic gas \citep[e.g.,][]{Weinberg1972}.

\subsection{Boundary Effects}
\label{sect:bound}

The tangled field must be confined within a certain volume (e.g., a
current sheet, see Figure \ref{fig:cartoon2}b), and the confinement
must be effective for a period much longer than the Alfven travel time
across the volume. What are the conditions for such confinement?
To answer this question we must consider the boundary region between a
tangled field and a smooth field. The tangled field is assumed to be
characterized by a high value of $|\alpha|$, and the smooth field
presumably has a much lower value of $| \alpha |$. To last a long
time, the magnetic field near the boundary must be approximately in
equilibrium (non-linear force-free field). The force-free condition
(\ref{eq:FFF}) implies that $\alpha$ is constant along field lines,
so there cannot be many field lines that pass from the smooth region
to the tangled region.  Therefore, one important condition for the
survival of the tangled field is that the two regions are nearly
disconnected from each other magnetically. Another requirement is that
the two regions are approximately in pressure balance.

To show that these conditions can be satisfied, we now consider a
simple model for the boundary region. The interface between the
tangled and smooth fields is approximated by a plane surface, here
taken to be the plane $x=0$ in Cartesian coordinates. The
above-mentioned condition on the lack of connectivity between the
smooth and the tangled fields requires $B_x (0,y,z)=0$ at $x=0$. The
tangled field in $x \ge 0$ is assumed to be a LFFF with a specified
value of $\alpha$. The solution of the LFFF equation is again written
as a superposition of planar modes. However, in the present case
the modes are grouped into pairs with closely related wave vectors
${\bf k}_n$ and ${\bf k}_n^\prime$, and with the same amplitude $B_n$
and phase $\beta_n$:
\begin{eqnarray}
{\bf B} ({\bf r}) & = & \sum_{n=1}^{N/2} B_n \left[ \hat{\bf e}_{1,n}
\cos ( {\bf k}_n \cdot {\bf r} + \beta_n )  - \hat{\bf e}_{2,n}
\sin ( {\bf k}_n \cdot {\bf r} + \beta_n ) \right. \nonumber
\\ 
 &  & \left. \cdots \,\,
- \hat{\bf e}_{1,n}^\prime \cos ( {\bf k}_n^\prime \cdot {\bf r} +
\beta_n ) + \hat{\bf e}_{2,n} \sin ( {\bf k}_n^\prime \cdot {\bf r} +
\beta_n ) \right] . \label{eq:sum2}
\end{eqnarray}
Here $N$ is the total number of modes, $\hat{\bf e}_{1,n}$ and
$\hat{\bf e}_{2,n}$ are defined in equations (\ref{eq:e1}) and
(\ref{eq:e2}), ${\bf k}_n^\prime \equiv \alpha
\hat{\bf e}_{3,n}^\prime$ is the modified wave vector, and the unit
vectors $\hat{\bf e}_{1,n}^\prime$ and $\hat{\bf e}_{3,n}^\prime$ are
defined by
\begin{eqnarray}
\hat{\bf e}_{1,n}^\prime & = & - \cos \theta_n  ( \cos \phi_n ~
\hat{\bf y} + \sin \phi_n ~ \hat{\bf z} ) - \sin \theta_n ~
\hat{\bf x} , \\
\hat{\bf e}_{3,n}^\prime & = & ~~ \sin \theta_n  ( \cos \phi_n ~
\hat{\bf y} + \sin \phi_n ~ \hat{\bf z} ) - \cos \theta_n ~
\hat{\bf x} .
\end{eqnarray}
Note that $\hat{\bf e}_{3,n}^\prime$ differs from $\hat{\bf e}_{3,n}$
only in the sign of the $x$-component, whereas
$\hat{\bf e}_{1,n}^\prime$ has the sign of the $y$- and $z$-components
reversed. Therefore, the unit vectors $[\hat{\bf e}_{1,n}^\prime,
\hat{\bf e}_{2,n}, \hat{\bf e}_{3,n}^\prime ]$ again form a
right-handed basis system. The magnetic field at the boundary ($x=0)$
is given by
\begin{displaymath}
{\bf B} (0,y,z) = 2 \sum_{n=1}^{N/2} B_n \cos \theta_n
( \cos \phi_n ~ \hat{\bf y} + \sin \phi_n ~ \hat{\bf z} ) 
\end{displaymath}
\begin{equation}
\times \,\,
\cos \left[ \alpha \sin \theta_n ( y \cos \phi_n + z \sin \phi_n )
+ \beta_n \right] , \label{eq:sum3}
\end{equation}
which satisfies $B_x (0,y,z) = 0$. Therefore, it is possible to
construct a tangled field that is disconnected from its surroundings.

We now consider the statistical average of the tensor ${\bf B}
{\bf B}$ at $x=0$. Averaging over phase angles $\beta_n$, we obtain
\begin{displaymath}
< {\bf B} {\bf B} >_\beta \, = 2 \sum_{n=1}^{N/2} B_n^2 \cos^2 \theta_n 
( \cos \phi_n ~ \hat{\bf y} + \sin \phi_n ~ \hat{\bf z} ) 
\end{displaymath}
\begin{equation}
\times \,\,
( \cos \phi_n ~ \hat{\bf y} + \sin \phi_n ~ \hat{\bf z} ) 
~~~ \mbox{at} ~ x=0,
\end{equation}
and further averaging over mode amplitudes and direction angles yields
\begin{equation}
< {\bf B} {\bf B} > = \frac{1}{6} B_0^2 
( \hat{\bf y} \hat{\bf y} + \hat{\bf z} \hat{\bf z} ) 
~~~ \mbox{at} ~ x=0 . \label{eq:tens0}
\end{equation}
Here we assume an isotropic distribution of direction angles, and we
use $< B_n^2 > = B_0^2 / N$, where $B_0$ is the r.m.s.~field strength
in the interior of the tangled field (see Section \ref{sect:volume}).
Note that at the boundary $< B^2 (0,y,z) > = B_0^2 /3$, while in the
interior $< B^2 > = B_0^2$, so the r.m.s.~field strength at the
boundary is reduced by a factor $1/\sqrt{3}$ compared to that in the
interior. The magnetic pressure at $x=0$ is given by
\begin{equation}
\frac{B_{ext}^2} {8 \pi} \approx \frac{< B^2 (0,y,z) >} {8 \pi} =
\frac{B_0^2} {24 \pi} = p_t , \label{eq:Pext}
\end{equation}
where $p_t$ is the average pressure in the interior of the tangled
field [see equation (\ref{eq:stress})]. Equation (\ref{eq:Pext}) shows
that it is possible to maintain pressure balance between the tangled
field and its surroundings.

\subsection{Tangled Field in a Cylinder}
\label{sect:cylinder}

\begin{figure}
\epsscale{1.10}
\plotone{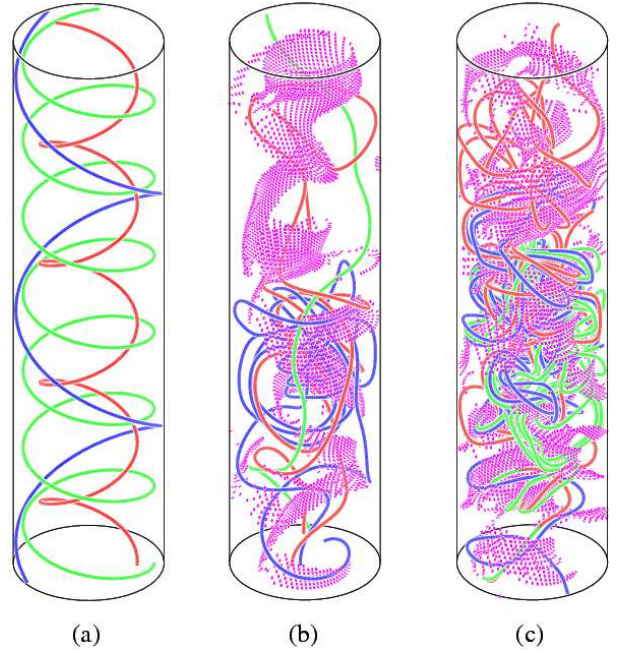}
\caption{Tangled magnetic fields obtained by superposition of modes of the
linear force-free field inside a cylinder, which represents a vertical
thread within a hedgerow prominence. The radial component of the
field $B_r$ vanishes at the cylinder wall. The three panels show
models with different values of $a \equiv |\alpha| R$, where $\alpha$
is the torsion parameter and $R$ is the cylinder radius: (a) $a =
3.0$; (b) $a = 4.5$ ; (c) $a = 6.0$. In case (a) the field has only a
single mode, but is cases (b) and (c) there are multiple modes ($N=4$
and $N=6$), some of which are non-axisymmetric, resulting in tangled
field lines. Each panel shows three field lines ({\it red}, {\it
green} and {\it blue} curves). In panels (b) and (c) the {\it
magenta} dots show dips in the field lines.}
\label{fig:tube}
\end{figure}

Here an infinitely long cylinder with radius $R$ is considered.
We adopt a cylindrical coordinate system $(r,\phi,z)$, and we assume
that the radial component of magnetic field vanishes at the cylinder
wall, $B_r (R,\phi,z) = 0$. In the Appendix we analyze the eigenmodes
of the LFFF equation in the domain $r \le R$ subject to the above
boundary condition. We find that this eigenvalue problem has a
discrete set of modes, and the number of modes depends on the
dimensionless parameter $a \equiv |\alpha| R$. Figure \ref{fig:tube}
shows the resulting magnetic fields for $a = 3.0$, 4.5 and 6.0. In the
first case only the axisymmetric (Lundquist) mode is present, so the
field lines are helical.  Assuming the cylinder axis is vertical,
there are no dips in the field lines. If cool plasma were to be
injected into such a structure, it would spiral down along the field
lines and quickly reach supersonic speeds. In contrast, for $a = 4.5$
and $a = 6.0$ there are multiple modes of the LFFF, and the random
superposition of these modes creates a tangled field with many dips
where prominence plasma can be supported. The field-line dips (i.e.,
sites where $B_z = 0$ and ${\bf B} \cdot \nabla B_z > 0$) are
indicated by {\it magenta} dots in the middle and right panels of
Figure~\ref{fig:tube}. We will return to this model in
Section~\ref{sect:thread}.

\section{Deviations from the Force-Free Condition}
\label{sect:forces}

The above models for a tangled field are purely force-free and do not
have any magnetic forces to support the prominence plasma against
gravity. To include such effects, we now consider the ``elastic''
properties of the tangled field, i.e., its response to external
forces. Specifically, the weight of the prominence causes the tangled
field to be compressed in the vertical direction, resulting in a
radially outward force on the plasma. Also, shearing motions may occur
within the tangled field as dense plasma moves downward and less dense
``plumes'' move upward \citep[e.g.,][]{Berger2008}. This results in
shear deformation of the tangled field and associated magnetic
stresses that counteract the plasma flows. In the following both of
these effects are considered in some detail.

\subsection{Compressional Effect}
\label{sect:compress}

We first consider the effects of gravity on a layer of tangled
magnetic field. The magnetostatic equation ($- \nabla p + \rho {\bf g}
+ {\bf F} = 0$) cannot be solved analytically for a tangled field,
so we make the following approximation:
\begin{equation}
{\bf B}^\prime ({\bf r}) \equiv \nabla \times \left[ \frac{1}{\alpha}
{\bf B} ({\bf r}) e^{-z/H_B} \right] = \left[ {\bf B} ({\bf r}) -
\frac{1}{\alpha H_B} \hat{\bf z} \times {\bf B} ({\bf r}) \right]
e^{-z/H_B} . \label{eq:compress}
\end{equation}
Here ${\bf B} ({\bf r})$ is the LFFF given by equation (\ref{eq:sum}),
and $H_B$ is the magnetic scale height of the modified field
(we assume $|\alpha| H_B > 2$). Note that $\nabla \cdot
{\bf B}^\prime = 0$ as required. This modified field ${\bf B}^\prime
({\bf r})$ is no longer force-free, but has the following statistical
properties:
\begin{eqnarray}
 & & < {\bf B}^\prime > = 0 , \label{eq:bavg1} \\
 & & < {\bf B}^\prime {\bf B}^\prime > = \onethird B_0^2 \left[
(1+\epsilon^2) (\hat{\bf x} \hat{\bf x} + \hat{\bf y} \hat{\bf y}) +
\hat{\bf z} \hat{\bf z} \right] e^{-2z/H_B} , \label{eq:tens2b} 
\end{eqnarray}
where $\epsilon \equiv (|\alpha| H_B)^{-1} < \onehalf$. Therefore, the
magnitude of the modified field drops off exponentially with height
$z$. Let $T_{ij}^\prime$ be the magnetic stress tensor of the modified
field. Taking its statistical average, we find for the nonzero
components of the stress tensor:
\begin{eqnarray}
 & & < T_{xx}^\prime > = < T_{yy}^\prime > = - \frac{B_0^2} {24 \pi}
~ e^{-2z/H_B} , \label{eq:Mxx} \\
 & & < T_{zz}^\prime > = - \frac{B_0^2} {24 \pi}
( 1 + 2 \epsilon^2 ) ~ e^{-2z/H_B} . \label{eq:Mzz}
\end{eqnarray}
Note that for $\epsilon < \onehalf$ the stress tensor is nearly
isotropic. The net force on the plasma is given by $F_i^\prime =
\partial T_{ij}^\prime / \partial x_j$, and the average force follows
from equations (\ref{eq:Mxx}) and (\ref{eq:Mzz}):
\begin{eqnarray}
< F_x^\prime > & = & < F_y^\prime > = 0 , \label{eq:Fxyp} \\
< F_z^\prime > & = & \frac{B_0^2}{12 \pi H_B} 
( 1 + 2 \epsilon^2 ) ~ e^{-2z/H_B} . \label{eq:Fzp}
\end{eqnarray}
Note that the average force acts in the positive $z$ direction, i.e.,
the magnetic force counteracts the force of gravity. In effect, the
plasma is being supported by the magnetic pressure of the tangled field.
The tangled field acts like a hot gas that has a significant pressure
but no mass. The {\it average} density of the plasma that can be
supported by the tangled field is given by
\begin{equation}
\rho_{avg} (z) = \frac{< F_z^\prime >}{g} = \frac{B_0^2}
{12 \pi g H_B} ~ ( 1 + 2 \epsilon^2 ) ~ e^{-2z/H_B} ,
\label{eq:rhoa1}
\end{equation}
where $g$ is the acceleration of gravity.

The horizontal components of Lorentz force, $F_x^\prime$ and
$F_y^\prime$, do not vanish for any particular realization of the
tangled field, and cannot be written as the gradients of a scalar
pressure $p$. The reason is that expression (\ref{eq:compress}) is not
an exact solution of the magnetostatic equilibrium equation. However,
equation (\ref{eq:Fxyp}) shows that the horizontal forces vanish when
averaged over the fluctuations of the isotropic tangled
field. Therefore, expression (\ref{eq:compress}) is thought to give a
good approximation for the effects of gravity on the tangled field.

We now apply the above model to the vertical threads observed in
hedgerow prominences. To explain the observed heights of such
prominences, we require that the magnetic scale height $H_B$ is at
least 100 Mm. The size $\ell$ of the magnetic tangles is assumed to
be in the range 0.1--1 Mm, so $\epsilon = \ell / H_B \ll 1$.
For $B_0 = 10$ G and $H_B = 100$ Mm, we find $\rho_{avg}
\approx 10^{-14}$ $\rm g ~ cm^{-3}$, which corresponds to an average
(total) hydrogen density $n_{H,avg} \approx 5 \times 10^9$ $\rm
cm^{-3}$. This is only about 0.05 times the density $\rho_0 \approx
2 \times 10^{-13}$ $\rm g ~ cm^{-3}$ or $n_H \approx 10^{11}$
$\rm cm^{-3}$ typically observed in hedgerow prominences \citep[][]
{Engvold1976, Engvold1980, Hirayama1986}. This comparison shows
that the pressure of the tangled field {\it inside} a prominence
thread is not sufficient to support the weight of the prominence
plasma. To support the plasma with tangled fields, we need to take
into account the magnetic coupling between the vertical thread and
its surroudings. Such coupling is neglected in the above
plane-parallel model.

\subsection{Shear Stress Effect}
\label{sect:shear}

We now assume that the tangled field pervades not only the observed
vertical threads but also their local surroundings. The density in the
surroundings is less than that in the threads, so the force of gravity
is also much lower. This difference in gravitational forces leads to
vertical motions (downflows in the dense threads, upflows in the
tenuous surroundings) that create magnetic stresses in the tangled
field. The magnetic coupling between the prominence and its
surroundings causes the weight of the dense prominence to be
distributed over a wider area. In effect, the prominence plasma is
being supported by the radial gradient of the magnetic pressure of the
tangled field over this larger area. In the following we estimate the
magnetic stresses and vertical displacements resulting from these
forces.

The tangled field is modeled either as a vertical slab with half-width
$R$, or as a vertical cylinder with radius $R$. The prominence is
located at the center of this slab or cylinder, and has a half-width
or radius $r_0 < R$. Then the average density in the tangled field
region is given by
\begin{equation}
\rho_{avg} \approx \rho_0 \left( \frac{r_0}{R} \right)^n ,
\label{eq:rhoa2}
\end{equation}
where $\rho_0$ is the density of the prominence, and we neglect
the mass of the surroundings. The exponent $n=1$ for the slab model
or $n=2$ for the cylindrical model. As discussed in Section
\ref{sect:compress}, observations of hedgerow prominences indicate
$\rho_0 \approx 2 \times 10^{-13}$ $\rm g ~ cm^{-3}$ \citep[e.g.,][]
{Engvold1976}, and to explain the observed heights of such prominences
with $B_0 = 10$ G, we require $\rho_{avg} / \rho_0 < 0.05$. According
to equation (\ref{eq:rhoa2}), this implies $R/r_0 \ge 20$ for the slab
model, or $R/r_0 \ge 4.5$ for the cylindrical model. The observed
threads have widths down to about 500 km \citep[][] {Engvold1976},
which corresponds to $r_0 \approx 250$ km. Therefore, the magnetic
coupling by the tangled field must extend to a surrounding distance
of at least 5 Mm for the slab model, or 1.1 Mm for the cylindrical
model. More generally, equations (\ref{eq:rhoa1}) and (\ref{eq:rhoa2})
yield the following expression for the magnetic scale height of
tangled field:
\begin{equation}
H_B \approx \frac{B_0^2} {12 \pi g \rho_0} \left( \frac{R}{r_0}
\right)^n .
\end{equation}
Therefore, the maximum height of the prominence depends strongly
on the magnetic field strength $B_0$.

According to the present model, magnetic stress builds up in the
tangled field as a result of the difference in gravitational forces
between the thread and its surroundings. Can the field support such
shear stress? To answer this question we examine the effect of
vertical displacements on the tangled field. For simplicity we neglect
the mean vertical force given by equation (\ref{eq:Fzp}), and we focus
on {\it relative} displacements. Let ${\bf r}^\prime$ be the new
position of a fluid parcel originally located at position ${\bf
r}$. In the limit of a perfectly conducting plasma, the deformed
field ${\bf B}^{\prime}$ at the new position ${\bf r}^\prime$ is given
by
\begin{equation}
B_i^\prime = \frac{B_j} {J} \frac{\partial x_i^\prime}
{\partial x_j} ,
\end{equation}
where ${\bf B} ({\bf r})$ is the original field, and $J$ is the
Jacobian of the transformation \citep[e.g.,][] {Priest1982}. We
assume
\begin{equation}
x^\prime = x, ~~~~ y^\prime = y, ~~~~ z^\prime = z + h(x,y) ,
\end{equation}
which yields
\begin{equation}
B_x^\prime = B_x , ~~~~ B_y^\prime = B_y , ~~~~
B_z^\prime = B_z + B_x \frac{\partial h}{\partial x} + 
B_y \frac{\partial h}{\partial y} ,
\end{equation}
where $h(x,y)$ is the vertical displacement. The original field
${\bf B} ({\bf r})$ is assumed to be a realization of the isotropic
tangled field given by equation (\ref{eq:sum}). Using equation
(\ref{eq:tens2}), we obtain
\begin{eqnarray}
 & & < {\bf B}^\prime > = 0 , \label{eq:bavg3} \\
 & & < {\bf B}^\prime {\bf B}^\prime > = \onethird B_0^2 \left[
\hat{\bf x} \hat{\bf x} + \hat{\bf y} \hat{\bf y}
+ f \hat{\bf z} \hat{\bf z}
+ \frac{\partial h}{\partial x} \hat{\bf x} \hat{\bf z}
+ \frac{\partial h}{\partial y} \hat{\bf y} \hat{\bf z} \right] ,
\label{eq:tens3} 
\end{eqnarray}
where
\begin{equation}
f(x,y) \equiv 1 + \left( \frac{\partial h} {\partial x} \right)^2 +
\left( \frac{\partial h} {\partial y} \right)^2 .
\end{equation}
This yields the following expressions for the off-diagonal components
of the stress tensor:
\begin{equation}
< T_{xz}^{\prime} > = \frac{B_0^2} {12 \pi} \frac{\partial h}
{\partial x} , ~~~~
< T_{yz}^{\prime} > = \frac{B_0^2} {12 \pi} \frac{\partial h}
{\partial y} .
\end{equation}
The Lorentz force is given by $F_i^{\prime} = \partial
T_{ij}^{\prime} / \partial x_j$, and since $< T_{zz}^{\prime} >$
is independent of $z$, the average vertical force is given by
\begin{equation}
< F_z^{\prime} > = \frac{B_0^2} {12 \pi} \left(
\frac{\partial^2 h} {\partial x^2} + 
\frac{\partial^2 h} {\partial y^2} \right)
= g \Delta \rho(x,y) ,
\label{eq:Fzpp}
\end{equation}
where $\Delta \rho (x,y)$ is the density perturbation ($\Delta \rho
\equiv \rho - \rho_{avg}$). Using this equation, we can determine the
vertical displacement $h(x,y)$ for a given density variation $\Delta
\rho (x,y)$. In the following subsections we solve the above equation
for the slab and cylinder models.

\subsubsection{Slab Model}
\label{sect:slab}

\begin{figure}
\epsscale{1.09}
\plotone{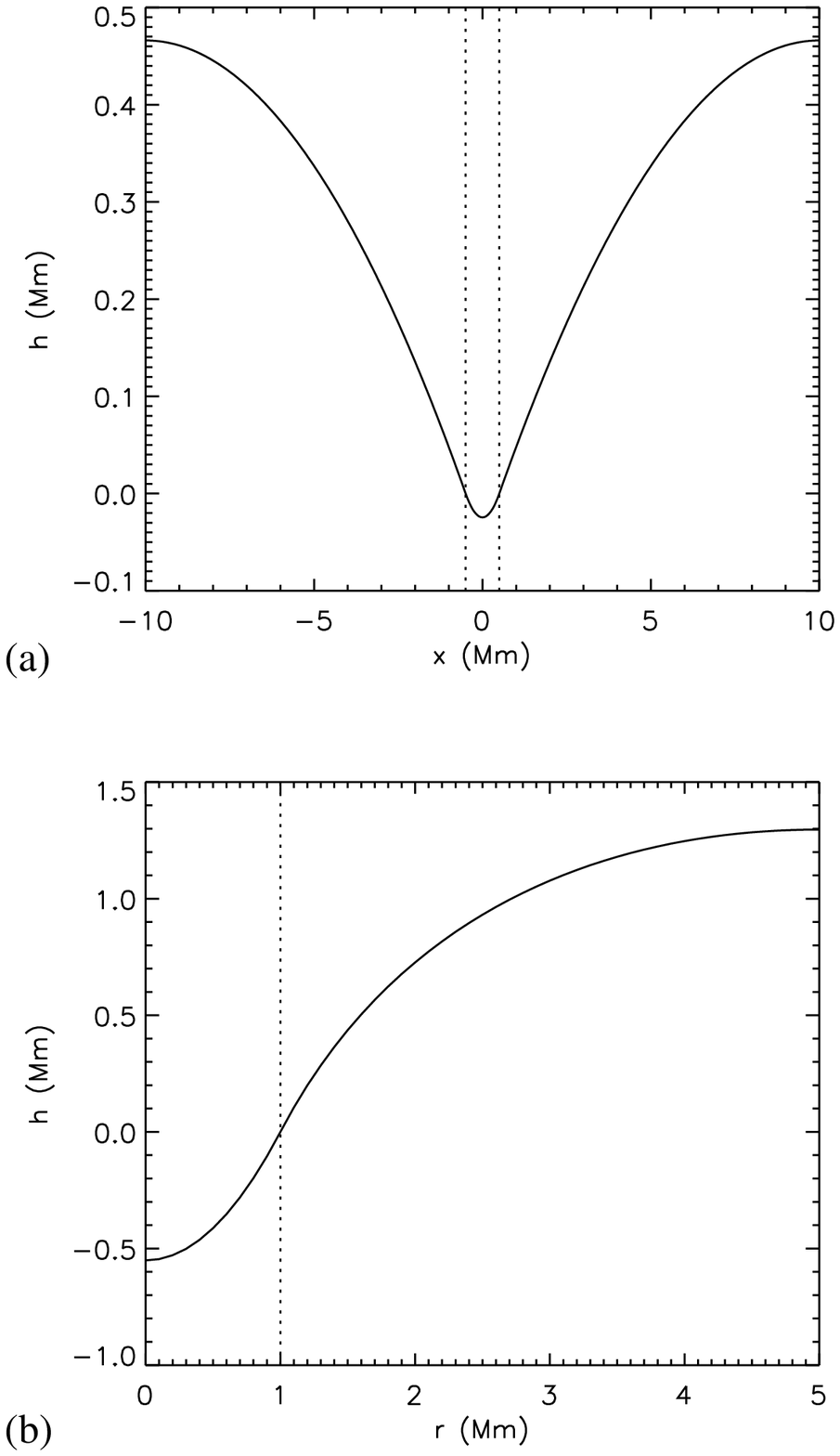}
\caption{Vertical displacements in the tangled magnetic field
supporting a hedgerow prominence. (a) Slab model with $R = 10$ Mm,
$r_0 = 0.5$ Mm, and $B_0 = 10$ G. The vertical displacement $h(x)$ is
plotted as function of position $x$ perpendicular to the vertical
slab. The location of the prominence is indicated by the dotted lines.
(b) Cylindrical model of a prominence thread with $R = 5$ Mm,
$r_0 = 1$ Mm, and $B_0 = 3$ G. In both cases the prominence density
$\rho_0 = 2 \times 10^{-13}$ $\rm g ~ cm^{-3}$.}
\label{fig:shear}
\end{figure}

We first consider a slab with infinite extent in the $+y$, $-y$ and
$+z$ directions. The coordinate $x$ perpendicular to the slab is in
the range $-R < x < R$, where $R$ is the half-width of the sheet in
which the tangled field is embedded (see Figure~\ref{fig:cartoon2}b).
Then the density perturbation is given by
\begin{equation}
\Delta \rho (x) = \left\{ \begin{array}{ll}
+ \rho_0 [1-(r_0/R)] & \mbox{for} ~ |x| < r_0 , \\ 
- \rho_0 r_0/R & \mbox{otherwise} . \label{eq:drho1}
\end{array} \right.
\end{equation}
Inserting this expression into equation (\ref{eq:Fzpp}) and solving
for the vertical displacement, we obtain
\begin{equation}
h(x) = \left\{ \begin{array}{ll}
C [1-(r_0/R)] (x^2 - r_0^2)  & \mbox{for} ~ |x| < r_0 , \\
C (r_0/R) [ (R-r_0)^2 - (R-|x|)^2 ] & \mbox{otherwise} ,
\end{array} \right.
\end{equation}
where $C \equiv 6 \pi g \rho_0 B_0^{-2}$.  Here we applied no-stress
boundary conditions ($dh/dx = 0$) at $x = \pm R$. Note that $h(x)$
and its derivative are continuous at the edges of the prominence
($x = \pm r_0$). The relative displacement across the tangled field is
given by
\begin{equation}
\Delta h \equiv h(R) - h(0) = \frac{6 \pi g \rho_0} {B_0^2}
r_0 (R-r_0) .  \label{eq:dh1}
\end{equation}
Figure \ref{fig:shear}a shows the function $h(x)$ for $r_0 = 0.5$ Mm,
$R = 10$ Mm, $B_0 = 10$ G and $\rho_0 = 2 \times 10^{-13}$
$\rm g ~ cm^{-3}$, so that $\Delta h = 0.491$ Mm. Note that
a relatively small deformation of the tangled field ($\Delta h \ll R$)
is sufficient to redistribute the gravitational forces over the full
width of the tangled field. However, $\Delta h$ is larger than the
pressure scale height of the prominence plasma ($H_p \approx 0.2$ Mm).
Therefore, the deformation of the tangled field in the neighborhood of
the prominence may have a significant effect on the distribution of
the prominence plasma. This issue will be further discussed in
Section~\ref{sect:thread}.

For comparison of the tangled field slab model with the flux rope
model (Figure \ref{fig:cartoon2}), we define the average {\it sag
angle} $\theta$ of the prominence relative to its surroundings:
\begin{equation}
\tan \theta \equiv \frac{\Delta h}{R} \approx \frac{6 \pi g \rho_0
r_0} {B_0^2} ,
\end{equation}
where $B_0$ is the r.m.s.~field strength of the tangled field, and we
assumed $r_0 \ll R$. This expression is similar to that derived for the
Kippenhahn-Schl\"{u}ter model:
\begin{equation}
\tan \theta = \frac{4 \pi g \rho_0 r_0} {B_x^2} , \label{eq:KS}
\end{equation}
where $B_x$ is the magnetic field through the mid-plane of the
prominence \citep[][] {Kippenhahn1957}. Equation (\ref{eq:KS})
describes the angle of the field lines in the flux rope model shown
in Figure \ref{fig:cartoon2}a. Therefore, the flux-rope and tangled
field models are similar in their ability to explain the magnetic
support of the prominence plasma, provided the half-width $R$ of the
tangled field region is similar to the radius $R$ of the flux rope.

\subsubsection{Cylindrical Model}
\label{sect:cyl}

We now consider a cylindrical model for a prominence thread with $r$
the distance from the (vertical) thread axis. The density perturbation
is given by
\begin{equation}
\Delta \rho (r) = \left\{ \begin{array}{ll}
+ \rho_0 [1-(r_0/R)^2] & \mbox{for} ~ r < r_0 , \\ 
- \rho_0 (r_0/R)^2 & \mbox{for} ~ r_0 < r < R . \label{eq:drho2}
\end{array} \right.
\end{equation}
The vertical displacement is obtained by solving equation
(\ref{eq:Fzpp}), which yields
\begin{equation}
h(r) = \left\{ \begin{array}{ll}
h_0 [(r/r_0)^2-1]  & \mbox{for} ~ r < r_0 , \\
h_0 [2 R^2 \ln (r/r_0) -r^2 + r_0^2 ]/(R^2-r_0^2) &
\mbox{for} ~ r_0 < r < R , \end{array} \right. \label{eq:hr}
\end{equation}
where
\begin{equation}
h_0 \equiv \frac{3 \pi} {B_0^2} g \rho_0 r_0^2
\left( 1 - \frac{r_0^2}{R^2} \right) ,
\end{equation}
and we applied no-stress boundary conditions at $r = R$. Then the
total displacement across the tangled field is
\begin{equation}
\Delta h \equiv h(R) - h(0) = \frac{6 \pi g \rho_0} {B_0^2}
r_0^2 \ln \left( \frac{R}{r_0} \right) .  \label{eq:dh2}
\end{equation}
Figure \ref{fig:shear}b shows the vertical displacement $h(r)$ for
$R = 5$ Mm, $r_0 = 1$ Mm, $B_0 = 3$ G and $\rho_0 = 2 \times
10^{-13}$ $\rm g ~ cm^{-3}$, which yields $h_0 = 0.551$ Mm and
$\Delta h = 1.847$ Mm.
In this case $\Delta h$ is significantly larger than the pressure
scale height ($H_p \approx 0.2$ Mm), mainly because of the lower
field strength compared to the case shown in Figure \ref{fig:shear}a.
In Section~\ref{sect:thread} we consider the effect of such
deformation on the field-line dips, and on the spatial distribution of
the prominence plasma.

The above analyses only provide an rough estimate for the density of
prominence plasma that can be supported by the tangled field. The
actual density distribution $\rho ({\bf r},t)$ is likely to be much
more complex for several reasons. First, plasma will tend to collect
at the dips of the field lines, so the density will vary on the
spatial scale $\ell$ of the tangled field and on the scale of $H_p$;
this effect will be considered in more detail in Section
\ref{sect:thread}. Second, the density will vary with time because
there are flows along the field lines (see Section \ref{sect:flows})
and these flows are likely to be non-steady. Also, the magnetic
structure is not fixed and will continually evolve as dipped field
lines are distorted by the weight of the prominence plasma.
To predict the actual density will require numerical simulations of
the interaction of tangled fields with prominence plasma, which is
beyond the scope of the present paper.

\section{Formation of Vertical Threads by Rayleigh-Taylor
Instability}
\label{sect:RT}

According to the present theory, hedgerow prominences are supported
by the pressure of a tangled magnetic field, which acts like a tenuous
gas and is naturally buoyant.  It is well known that a tenuous medium
supporting a dense medium is subject to Rayleigh-Taylor (RT)
instability \citep[][] {Chandra1961}. Therefore, we suggest that the
observed vertical threads may be a consequence of RT instability
acting on the tangled field and the plasma contained within it. As
cool plasma collects in certain regions of the tangled field, the
weight of the plasma deforms the surrounding field, which causes even
more plasma to flow into these regions.

A detailed analysis of the formation of prominence threads by RT
instability is complicated by the fact that we presently do not
understand how a tangled field responds to shear deformation. In
Section \ref{sect:shear} we estimated the relative vertical
displacement $\Delta h$ of the prominence plasma assuming no
reconnection occurs during the deformation of the magnetic field by
gravity forces [see equations (\ref{eq:dh1}) and (\ref{eq:dh2})].  In
this case the tangled field behaves as an ``elastic'' medium with
magnetic forces proportional to the displacement. However, it is not
clear that this approximation is valid. High-resolution observations
of prominences indicate that the dense threads move downward relative
to their more tenuous surroundings with velocities of the order of
10-30 $\rm km ~ s^{-1}$ \citep[e.g.,][] {Berger2008, Chae2008}. If the
threads and their surroundings are indeed coupled via tangled fields,
these relative motions imply that the field is continually being
stretched in the vertical direction. Therefore, the shear stress
continually increases with time, unless there is internal reconnection
that causes the shear stress to be reduced.

We speculate that tangled fields have a tendency to relax to the LFFF
via internal reconnection. A similar relaxation processes occurs in
the reversed field pinch and other laboratory plasma physics devices
\citep[][] {Taylor1974}. Therefore, the long-term evolution of
prominence threads likely involves small-scale reconnection within the
tangled field. The tangled field may behave more like a ``plastic''
medium that is irreversibly deformed when subjected to shear stress. 
Such plasticity makes it possible to understand how the dense threads
can move downward relative their the surroundings at a small but
constant speed. These flows significantly deform the tangled field,
but the field is nevertheless able to support the plasma against
gravity. A detailed analysis of reconnection in tangled fields and
its effect on the prominence plasma is beyond the scope of the
present paper. 

The observed vertical structures likely reflect the non-linear
development of the RT instability in hedgerow prominences. To
establish a vertical column of mass resembling a prominence thread
will likely require vertical motions over a significant height range
(tens of Mm). Starting from a homogeneous density distribution, it may
take several hours for the threads to form by RT instability.

\section{Model for a Prominence Thread}
\label{sect:thread}

We now construct a model for the density distribution in a fully
formed (vertical) prominence thread supported by a tangled field.
It is assumed that the RT instability has produced a vertical thread
that is clearly separated from the rest of the prominence plasma.
Therefore, only a single thread and its local surroundings are
considered, and the tangled field is assumed to be contained in a
vertical cylinder with radius $R = 5$ Mm. As discussed in Section
\ref{sect:flows}, there will in general be mass flows along the
tangled field lines, but for the purpose of the present model we
neglect such flows and we assume that the plasma is in hydrostatic
equilibrium along the field lines.

To construct the density model, we first compute a particular
realization of the LFFF with $\alpha R = 9$ (see Appendix for
details). To account for the weight of the prominence plasma, this
field is further deformed as described by equation (\ref{eq:hr}).
The deformation parameters are $r_0 = 1.25$ Mm and $h_0 = 1.5$ Mm,
which yields $\Delta h = 4.44$ Mm, somewhat larger than the values
used in Figure \ref{fig:shear}b. As shown in Section
\ref{sect:compress}, the weight of the prominence plasma causes the
strength of the tangled field to decrease with height [see equation
(\ref{eq:compress})], but for simplicity this gradient is neglected
here. The pressure scale height is assumed to be constant, $H_p =
0.2$ Mm, which corresponds to a temperature of about 8000 K, typical
for H$\alpha$ emitting plasma in prominences.

\begin{figure}
\epsscale{1.10}
\plotone{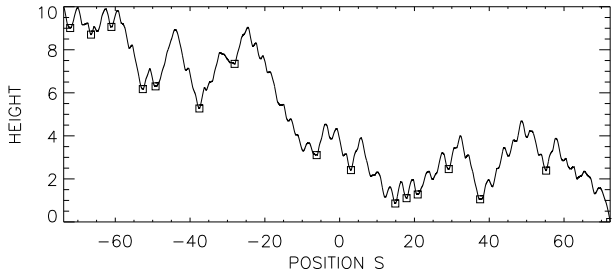}
\caption{Height $z(s)$ as function of position $s$ along a field line
in the cylindrical tangled field model with $a = 9.0$. Lengths are in
units of the cylinder radius $R$.}
\label{fig:dips}
\end{figure}

We introduce cartesian coordinates $(x,y,z)$ with the $z$ axis along
the cylinder axis; the $x$ and $y$ coordinates are in the range
$[-R,+R]$, and the height $z$ is in the range $[0,10R]$. The density
$\rho (x,y,z)$ in this volume is computed on a grid with $200 \times
200 \times 1000$ grid points, using the following method. We randomly
select a large number of points within the cylinder and trace out the
field lines that pass through these points. For each field line we
plot the height $z(s)$ as function of position $s$ along the field
line, and we find the peaks and dips in the field line. For each dip
we find the two neighboring peaks ($s_1$ and $s_2$) and we determine
the depth $\Delta z$ of the valley between these peaks. We then
iteratively remove shallow dips
with $\Delta z < 3 H_p$ by concatenating neighboring sections.
Figure \ref{fig:dips} shows an example for one particular field line;
the remaining dips with $\Delta z > 3 H_p$ are indicated by
squares. We then compute the density $\rho (s)$ in each interval,
assuming hydrostatic equilibrium along the field line [see equation
(\ref{eq:rho})]. Since we are interested only in {\it relative}
densities, we set $\rho_{dip} = 1$, the same for all dips on all field
lines. Finally, we distribute the density $\rho (s)$ onto the 3D grid
by finding the grid points that lie closest to the path of the field
line. This process is repeated for 8000 field lines to obtain the
density $\rho (x,y,z)$ throughout the 3D grid.

\begin{figure}
\epsscale{1.01}
\plotone{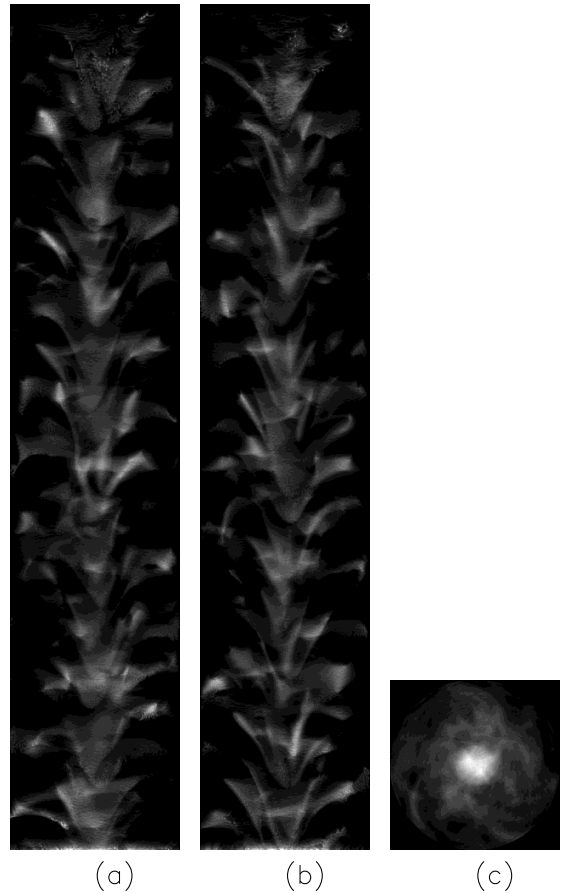}
\caption{Simulated density distribution of a prominence thread.
The thread is supported by a tangled magnetic field with $R = 5$ Mm
and $\alpha R = 9$. The field is distorted by the weight of the
prominence plasma ($r_0 = 1.25$ Mm, $h_0 = 1.5$ Mm).
The plasma is concentrated at the dips of the tangled field lines.
The density distribution along the field lines assumes hydrostatic
equilibrium with $H_p = 0.2$ Mm. The three panels show the column
density distributions as seen from the $x$, $y$ and $z$ directions,
respectively. Note that multiple sheets of plasma are superposed along
the line of sight.}
\label{fig:mass_plot}
\end{figure}

Figure \ref{fig:mass_plot} shows the resulting density distribution.
The three panels show different projections obtained by integration
in the $x$, $y$ and $z$ directions, respectively [for example,
Figure~\ref{fig:mass_plot}a shows $\int \rho (x,y,z) dx$].
Note that the plasma is concentrated in the central part of the
cylinder; this is due to the deformation of the magnetic field
described by the vertical displacement $h(r)$, which changes the
distribution of the field-line dips compared to that in the LFFF. The
plasma is highly inhomogenous ($\rho_{max} \approx 50 \rho_{avg}$),
and is distributed in sheets corresponding to surfaces of field-line
dips. In some regions there are multiple sheets along the line of
sight (LOS). This is consistent with observations of the H~I Lyman
lines in solar prominences, which indicate multiple threads along the
LOS \citep[][] {Orrall1980, Gunar2007}.

The above model is quasi-static and does not take into account the
expected mass flows along the field lines (Section \ref{sect:flows}),
nor the dynamical processes by which the threads are formed (Section
\ref{sect:RT}). Constructing a more realistic 3D dynamical model of a
prominence thread supported by tangled fields is beyond the scope of
the present paper. However, our main conclusion that the density
distribution within the thread is highly inhomogeneous is likely to be
valid also in the dynamical case.

\section{Summary and Discussion}
\label{sect:discussion}

We propose that hedgerow prominences are supported by magnetic
fields that are ``tangled'' on a spatial scale of 1 Mm or less.
A key feature of the model is that the plasma is approximately in
magnetostatic balance, therefore, the model is different from earlier
models in which the plasma is supported by MHD-wave pressure
\citep[e.g.,][] {Jensen1983, Jensen1986, Pecseli2000}. In the present
case the perturbations of the magnetic field do not propagate along
the field lines. The tangled field is located within a large-scale
current sheet standing vertically above the PIL (Figure
\ref{fig:cartoon2}b), and is not magnetically connected to the
photosphere on either side of the PIL. Such tangled fields may be
formed by flux emergence followed by magnetic reconnection in the low
corona.

In this paper we use a variety of methods to explore the interactions
of prominence plasma with tangled fields. In Section \ref{sect:flows}
a simple model for the downward flow of plasma along the distorted
field lines was developed.  It was shown that such flows naturally
develop standing shocks where the gravitational energy of the plasma
is converted into heat; this may be important for understanding the
heating of prominence plasmas. The average flow velocity is less than
the sound speed, indicating that the tangled field is able to support
the prominence plasma against gravity.

In Section \ref{sect:LFFF} linear force-free models of tangled fields
were constructed. Tangled fields can be described as a superposition
of planar modes. We studied the statistical properties of such fields,
and found random-walk behavior of the field lines when the number of
modes is sufficiently large ($N > 10$). To produce tangling of the
field lines on a scale of 1 Mm or less, as required for our model of
hedgerow prominences, we need $| \alpha | > 1$ $\rm Mm^{-1}$, much
larger than the values typically found in measurements of photospheric
vector fields. Therefore, according to the present model, hedgerow
prominences are embedded in magnetic fields with high magnetic
helicity density. We also considered the conditions for confinement of
a tangled field by the surrounding smooth fields, and showed that
despite the high helicity density tangled fields can be in pressure
balance with their surroundings. 

In Section \ref{sect:forces} the ``elastic'' properties of a tangled
field were considered, i.e., their linear response to gravitational
forces assuming ideal MHD. We found that the weight of the prominence
plasma can be supported by the nearly isotropic magnetic pressure of
the tangled field. The tangled field pervades not only the observed
vertical threads, but also their local surroundings. The magnetic
coupling between the threads and their surroundings is quite strong:
vertical displacements of only 0.5--2 Mm are sufficient to counteract
the shear stress resulting from the different gravitational forces.
In effect, the weight of a dense thread is distributed over an area
that is larger than the cross-sectional area of the thread. As
discussed in Section~\ref{sect:shear}, the observed densities in
prominence threads ($n_H \sim 10^{11}$ $\rm cm^{-3}$) can be supported
by tangled fields with field strengths in the range 3--15 G.

In Section \ref{sect:RT} we proposed that the observed vertical
structures in hedgerow prominences are a consequence of
Rayleigh-Taylor (RT) instability acting on the tangled field and the
plasma contained within it.  The tangled field acts like a hot,
tenuous gas and is naturally buoyant; its support of the dense
prominence plasma is likely to be unstable to flows that separate the
gas into dense and less-dense columns.  The observed vertical
structures likely reflect the non-linear development of the RT
instability. A detailed analysis of this instability is complicated by
the fact that it requires internal reconnection to occur within the
tangled field, and it is unclear how rapidly such reconnection can
proceed. Therefore, it is difficult to predict the vertical velocities
in prominence threads. Clearly, more advanced numerical simulations of
the interaction of tangled fields with prominence plasma are needed.

Finally, in Section \ref{sect:thread} we simulated the density
distribution in a prominence thread, using a cylindrical model for the
tangled field and its deformation by gravitational forces. The results
indicate that the threads have an intricate fine-scale structure.
Multiple structures are superposed along the LOS, consistent with
observations of the H~I Lyman lines \citep[][]{Orrall1980, Gunar2007}.
While our model does not take into account any dynamical processes,
the main conclusion that the density distribution is highly
inhomogeneous is likely to be valid also in the dynamical case.

\acknowledgements
The images shown in Figures \ref{fig:proms}a and \ref{fig:proms}b were
obtained at the Big Bear Solar Observatory, which is operated by the
New Jersey Institute of Technology. The image shown in Figure
\ref{fig:proms}c was obtained at the Dutch Open Telescope, which is
operated by Utrecht University in the Observatorio del Roque de los
Muchachos of the Instituto de Astrof\'{i}sica de Canarias, La Palma,
Spain.

\appendix

\section{Tangled Fields in a Cylinder}
\label{cylinder}

\begin{figure}
\epsscale{0.45}
\plotone{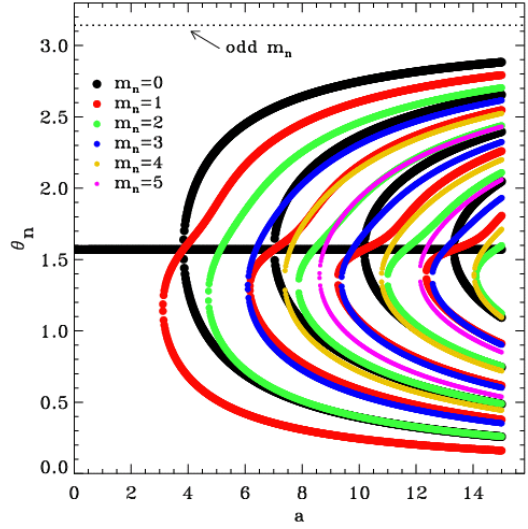}
\caption{Modes of the tangled field in a cylinder (see text for
details).}
\label{fig:modes}
\end{figure}

According to the model presented in this paper, the vertical threads
in hedgerow prominences are supported by tangled magnetic fields that
pervade the dense threads and their local surroundings. In this
section we construct a cylindrical model of this tangled field.
We use a cylindrical coordinate system $(r,\phi,z)$ with $r$ the
distance from the cylinder axis ($r \le R$ where $R$ is the cylinder
radius), $\phi$ the azimuthal angle, and $z$ the height along the
axis. We assume that the radial component of the field vanishes at the
cylinder wall, $B_r (R,\phi,z) = 0$, so the field lines are confined
to the interior of the cylinder. It follows that the axial magnetic
flux must be constant along the cylinder:
\begin{equation}
\Phi \equiv \int_0^R \int_0^{2 \pi} B_z(r,\phi,z) r ~dr ~d\phi =
\hbox{constant} . \label{eq:Phi1}
\end{equation}
In cylindrical coordinates the force-free condition reads:
\begin{eqnarray}
\frac{1}{r} \frac{\partial B_z} {\partial \phi} - 
\frac{\partial B_\phi} {\partial z} & = & \alpha B_r ,
\label{eq:f1} \\
\frac{\partial B_r} {\partial z} - 
\frac{\partial B_z} {\partial r} & = & \alpha B_\phi ,
\label{eq:f2} \\ 
\frac{1}{r} \frac{\partial} {\partial r} \left( r B_\phi \right) - 
\frac{1}{r} \frac{\partial B_r} {\partial \phi} & = & \alpha B_z .
\label{eq:f3}
\end{eqnarray}
We again take $\alpha$ to be constant (LFFF). The general solution of
the above equations with the boundary condition $B_r (R,\phi,z) = 0$
can be written as a superposition of discrete eigenmodes enumerated
by an index $n$:
\begin{eqnarray}
B_r    (r,\phi,z) & = & \sum_{n=1}^{N} \tilde{B}_n F_{r,n} (r)
\sin ( k_n z + m_n \phi + f_n ) , \label{eq:Br} \\ 
B_\phi (r,\phi,z) & = & \sum_{n=1}^{N} \tilde{B}_n F_{\phi,n} (r) 
\cos ( k_n z + m_n \phi + f_n ) , \label{eq:Bp} \\ 
B_z    (r,\phi,z) & = & \sum_{n=1}^{N} \tilde{B}_n F_{z,n} (r)
\cos ( k_n z + m_n \phi + f_n ) , \label{eq:Bz}
\end{eqnarray}
where $\tilde{B}_n$ is the mode amplitude; the functions $F_{r,n}(r)$,
$F_{\phi,n} (r)$ and $F_{z,n} (r)$ describe the radial dependence
of each mode; $k_n$ is the axial wavenumber; $m_n$ is the azimuthal
wavenumber (non-negative integer); and $f_n$ is the phase. Here $n=1$
refers to a fully symmetric mode with $k_1 = m_1 = f_1 = 0$ (also
known as the Lundquist mode), while $n=2, \cdots, N$ refers to modes
that have either a $\phi$ or $z$ dependence. The latter modes only
exist under certain conditions (see below). Inserting expressions
(\ref{eq:Br}), (\ref{eq:Bp}) and (\ref{eq:Bz}) into equations
(\ref{eq:f1}), (\ref{eq:f2}) and (\ref{eq:f3}), we find that the
radial dependencies can be expressed in terms of Bessel functions:
\begin{eqnarray}
F_{r,n} (r)    & = & - \frac{k_n}{q_n} J_m^\prime (q_n r)
- \frac{\alpha m_n} {q_n^2 r} J_m(q_n r) , \label{eq:Fr} \\ 
F_{\phi,n} (r) & = & - \frac{\alpha}{q_n} J_m^\prime (q_n r)
- \frac{k_n m_n} {q_n^2 r} J_m(q_n r) , \label{eq:Fp} \\ 
F_{z,n} (r)    & = & J_m(q_n r) , \label{eq:Fz}
\end{eqnarray}
where $q_n \equiv \sqrt{ \alpha^2 - k_n^2}$ is the radial wavenumber,
$J_m(x)$ is the Bessel function of order $m_n$, and $J_m^\prime (x)$
is its derivative. The boundary condition at the cylindrical wall
requires $F_{r,n} (R) = 0$ for all modes. Introducing a
parameter $\theta_n$ in the range $[0,\pi]$ such that $q_n = |\alpha|
\sin \theta_n$ and $k_n = \alpha \cos \theta_n$, we obtain the
following equation for $\theta_n$:
\begin{equation}
\cos \theta_n J_m^\prime (a \sin \theta_n) + \frac{m_n}
{a \sin \theta_n} J_m (a \sin \theta_n) = 0 , \label{eq:BC}
\end{equation}
where $a \equiv |\alpha| R$. The roots of this equation can be found
numerically for any given values of $a$ and $m$. Depending on the
value of $a$, the equation may have one or more solutions:
\begin{enumerate}
\item There always exists at least one solution, the axisymmetric mode
($m_1 = 0$) with $\theta_1 = \pi /2$. In this case $q_1 = |\alpha|$
and $k_1 = 0$, so this mode is invariant with respect to translation
along the $z$ axis (Lundquist mode).
\item
There may be additional axisymmetric modes that are not invariant
to translation. In the case $m_n = 0$ and $\theta_n \neq \pi/2$,
equation (\ref{eq:BC}) yields $a \sin \theta_n = x_i$, where $x_i$
is a root of $J_0^\prime(x) = 0$ ($x_1 = 3.832$, $x_2 = 7.016$,
$x_3 = 10.173$, etc.). Solutions exist only when $a > x_1$; for $a$
in the range $[x_1,x_2]$ there exists one such solution with
$\theta_n = \arcsin (x_1/a)$; for $a$ in the range $[x_2,x_3]$ there
exist two solution, etc. Since $J_0^\prime (a \sin \theta)$ is
symmetric with respect to $\theta = \pi /2$, each solution $\theta_n$
also has a second solution $\theta_n^\prime = \pi - \theta_n$, but
the magnetic structure of these solutions is the same, so we do not
count it as a separate mode.
\item
For higher values of $a$ there exist non-axisymmetric solutions
($m_n \geq 1$). The first modes with $m=1$ occur at $a = 3.112$;
these ``kink'' modes are apparently more easily excited than the
axisymmetric modes with $k \ne 0$. The first modes with $m=2$ occur at
$a = 4.708$.
\end{enumerate}
For a given value of $a$, we systematically find all roots of equation
(\ref{eq:BC}), starting with $m=0$ and then increasing $m$ until no
more roots are found. Figure \ref{fig:modes} shows $\theta_n$ as
function of $a$. Note that the number of modes $N$ increases with
$a$.

We now compute the axial flux and magnetic energy density. Inserting
equations (\ref{eq:Bz}) and (\ref{eq:Fz}) into equation
(\ref{eq:Phi1}), we find that only the Lundquist mode contributes to
the axial magnetic flux:
\begin{equation}
\Phi = 2 \pi \tilde{B}_1 \int_0^R J_0 ( |\alpha| r ) r ~dr =
2 \pi R \tilde{B}_1 |\alpha|^{-1} J_1 (a) . 
\label{eq:Phi2}
\end{equation}
where we used $J_0^\prime (q_n R) = 0$ for axisymmetric modes. The
mean magnetic energy density is defined by
\begin{equation}
E \equiv \frac{2}{R^2} \int_0^R \left\langle \frac{B^2} {8 \pi}
\right\rangle ~ r ~dr ,
\label{eq:E}
\end{equation}
where $< \cdots >$ denotes an average over $\phi$ and $z$. Let $n1$
and $n2$ denote two {\it different} modes, then averages of cross
products such as $<\cos (k_{n1} z + m_{n1} \phi + f_{n1})$
$\cos (k_{n2} z + m_{n2} \phi + f_{n2}) >$ vanish. Therefore, the
magnetic energy can be written as a sum over individual modes,
$E = \sum_{n=1}^N E_n$, where
\begin{eqnarray}
E_1 & = & \frac{\tilde{B}_1^2}{4 \pi R^2} \int_0^R
\left[ J_0^2 (|\alpha| r) + J_1^2 (|\alpha| r) \right] r ~dr ,
\label{eq:E1} \\
E_n & = & \frac{\tilde{B}_n^2}{8 \pi R^2} \int_0^R
\left[ F_{r,n}^2 (r) + F_{\phi,n}^2 (r) + F_{z,n}^2 (r) \right]
r ~dr ~~~~~ \mbox{for} ~ n > 1 . \label{eq:En}
\end{eqnarray}
For simplicity we consider {\it equipartition} tangled fields in which
the various modes $n$ of the LFFF have equal magnetic energy. This
implies $E_n = E/N$, the same for all modes (including the Lundquist
mode), which provides a relationship between the mode amplitudes
$\tilde{B}_n$. The phase angles $f_n$ of the non-axisymmetric modes
are assigned random values in the range $[0,2\pi]$. This results in a
tangled magnetic field ${\bf B} ({\bf r})$ described by equations
(\ref{eq:Br}), (\ref{eq:Bp}) and (\ref{eq:Bz}).

\end{document}